\def\wig#1{\mathrel{\hbox{\hbox to 0pt{%
          \lower.5ex\hbox{$\sim$}\hss}\raise.4ex\hbox{$#1$}}}}

\def\v1n{{\cal U}^N_1}

\def\sss{\scriptscriptstyle}
\def\phih2h2{\phi_{\sss {\rm H_2-H_2}}}
\def\phh2{\phi_{\sss {\rm H-H_2}}}

\def\Teff{T_{\rm eff}}
\def\sqr#1#2{{\vcenter{\vbox{\hrule height.#2pt 
  \hbox{\vrule width.#2pt height#1pt \kern#1pt 
  \vrule width.#2pt} 
  \hrule height.#2pt}}}}

\def\ref#1{\parindent=0pt\hangindent24pt\hangafter1
           \baselineskip=20pt{#1}\par }

\newcommand\beq{\begin{equation}}
\newcommand\eeq{\end{equation}}

 \documentstyle[aas2pp4]{article}

\begin{document}

\title{PURE HYDROGEN MODEL ATMOSPHERES FOR VERY COOL WHITE DWARFS}

\author{D. SAUMON AND S.B. JACOBSON}
\affil{Department of Physics and Astronomy, Vanderbilt University ,
   Nashville, TN 37235}

\begin{abstract}

Microlensing events observed in the line of sight toward the LMC indicate that
a significant fraction of the mass of the dark halo of the Galaxy is probably
composed of white dwarfs.  In addition, white dwarf sequences have now be observed 
in the HR diagrams of several globular clusters.  
Because of the unavailability of white dwarf atmospheres for 
$\Teff < 4000\,$K, cooling time scales for white dwarfs older than $\approx$10 Gyr are 
very uncertain.  Moreover, the identification of  a MACHO white dwarf population by 
direct observation depends on a knowledge of the colors and bolometric corrections
of very-cool white dwarfs.

In this paper we present the first detailed model  atmospheres and
spectra of very cool hydrogen white dwarfs  for $\Teff < 4000\,$K.  We include the
latest description of the opacities of hydrogen and  significantly, we 
introduce a non-ideal equation of state in the atmosphere calculation.  We find
that due to strong absorption from H$_2$ in the infrared, very old
white dwarfs are brightest in the $V$, $R$, and $I$ bands, and we confirm that they
become {\it bluer} in most color indices as they cool below $\Teff \approx 3500\,$K. 
\end{abstract}


\section{Introduction}

The MACHO and EROS microlensing experiments (\cite{ML1}; \cite{ML2}) have
detected a sizeable fraction of the dark halo of the Galaxy. Between $40\%-80\%$
of the halo may be composed of MACHO's with a probable mass of
$0.5^{+0.3}_{-0.2} M_{\odot}$ (\cite{ML1}).  Main sequence stars 
within this mass range would be relatively bright and
have been ruled out by direct observation (\cite{BFG}; \cite{VLMS2}; \cite{CM97}).
Since the mass function of disk white dwarfs is sharply peaked around 0.55$\,M_\odot$
(\cite{BSL92}) and that old white dwarfs are intrinsically very faint, the microlensing
events can be naturally explained by
a large population of old white dwarfs in the halo.
While suggestive, this interpretation of the microlensing events  is 
controversial (\cite{LF3}; \cite{SE2}; \cite{CSM}; \cite{SE3};
\cite{SE5}).   Ultimately, the nature of the MACHO's causing the observed 
events can only be settled by their direct detection.

The analysis of observational constraints on the population of halo white dwarfs rests
on models of the luminosity function which depends strongly on the physics of 
white dwarf cooling. 
(\cite{LF1}; \cite{LF2}; \cite{SE3}; \cite{CSM}). One of the most important sources
of uncertainty in the evolution of old white dwarfs is the physics of the atmosphere.   
In addition,  reliable synthetic spectra are required to
relate calculated cooling tracks to observations.
For $T_{\rm eff} <  4000\,$K such 
spectra have not been previously available. Extrapolations from higher $\Teff$
calculations, which have been commonly used in the analysis of
observational data, are poor approximations of the spectra of these stars. 
This is a result of the dramatic effect of collision-induced absorption (CIA) by 
molecular hydrogen on the emergent spectra of very cool hydrogen-rich
atmospheres (\cite{SB}; \cite{BSW}, hereafter BSW). 

Because pure helium atmospheres are much
more transparent than hydrogen-rich atmospheres, cool non-DA white dwarfs have comparatively
short cooling times and become extremely faint after 10 Gyr. They are not expected to 
be detectable at the characteristic ages and distances relevant to a halo population of
white dwarfs. Furthermore, hydrogen-rich white dwarfs are much more common 
than helium-rich white dwarfs.  The observation of an old halo population of
white dwarfs, in globular clusters or as a component of the dark halo, is primarily
concerned with hydrogen-rich white dwarfs. 

In this paper we provide an extension of the grid of models presented in
\cite{SB}  to pure hydrogen atmospheres in the $\Teff$ range 
1500$\,$K to 4000$\,$K  and gravity range $10^7\,$cm/s$^2$ to 10$^9\,$cm/s$^2$. 
The opacities are updated and a non-ideal equation of state (EOS) is introduced into
the calculation because
of the high densities encountered in the atmospheres of very cool
white dwarfs. We discuss these changes in the input physics  of the model calculation 
in \S2.  The resulting colors for very cool white dwarfs
are presented in \S3.  We conclude by discussing the consequences for studies
of old white dwarfs in the halo in \S4.

\section{Model Calculations}
  
The models presented here extend the calculation of zero-metallicity brown
dwarf atmospheres of Saumon et al. (1994) to higher gravities and the cool white dwarf
atmospheres grids of BSW and of \cite{BRL} (hereafter, BRL)
to lower $\Teff$.  Details of the atmosphere calculation
are provided in these references.  Briefly, we consider plane parallel atmospheres 
in LTE with constant gravity.  
Continuum opacity is provided by collision-induced absorption (CIA)  by H$_2$, Rayleigh
scattering by H$_2$ and H, H$^-$ bound-free and free-free, H$_2^-$ free-free, H$_2^+$ bound-free
and free-free.  Line opacity is negligible in these very cool atmospheres.
These atmospheres all become convective at depth and we use the mixing length
formalism to compute the convective flux.  As found by Saumon et al. (1994)
for brown dwarfs, convection is becomes extremely efficient deep in low-$\Teff$ models.
Consequently, the convection zone becomes nearly adiabatic at depth and the
structure of the atmosphere is insensitive to the choice of the mixing length parameter.

\subsection{Opacities}

For the range of $\Teff$ of interest in this paper, the dominant contributions to the
opacity are  H$_2$ CIA and H$_2$ Rayleigh scattering at temperatures below $\sim 3500 K$ 
and H$^-$ bound-free and free-free for higher temperatures. 
The H$_2$ CIA opacity is very strong in the infrared and largely determines the shape
of the spectral energy distribution (SED) 
for $\Teff \wig< 3000\,$K.  The H$_2$ CIA photo-absorption cross sections (\cite{ZB} and
references therein) have recently been revised by Borysow, J\o rgensen \& Zheng (1997).
The new CIA  cross sections are smaller above $\sim 1.2\,\mu$m and generally larger  
at shorter wavelengths. 
For  densities greater than 10$^{-2}\,$g/cm$^3$, collisions involving three H$_2$ molecules
are no longer negligible and the frequency-integrated CIA opacity 
then takes the form (\cite{KR}):

$$\kappa=\kappa_2 \rho^2 + \kappa_3 \rho^3 ,$$
where the ratio $\kappa_3/\kappa_2$ has been measured experimentally (\cite{K3}).

The H$^-$ ion is an important absorber when $T \wig> 3500\,$K and its
abundance is closely tied to the abundance of H$_3^+$
which is the main positive charge carrier in these models. The H$_3^+$ ion is
not a significant source of opacity but the structure and spectrum is quite sensitive to the
partition function of H$_3^+$ used in the calculation of the chemical equilibrium.
We have used the most recent determination of Neale \& Tennyson (1995).  
The principal source of uncertainty in our models lies in the use of an H$^-$ opacity
calculated for an isolated ion.  In the atmospheres considered here, the density
can become large enough to expect the bf and ff cross-sections of H$^-$ to differ
from their zero-density value.  To our knowledge, no one has attempted to compute
temperature- and density-dependent opacities for H$^-$ as a trace species in a dense
fluid of atoms and molecules.

BRL present observational evidence for an undetermined continuum opacity source affecting
the flux at $B$ and at shorter wavelengths.  They tentatively attribute this opacity
to bound-free transitions from the $n=1$ level of the hydrogen atom to higher levels which
are broadened into a pseudo-continuum by pressure effects.  We have not attempted to
model this pseudo-continuum source of opacity.  The importance of this opacity must
decrease very rapidly for $\Teff \wig< 4000\,$K as atoms recombine to form molecules.

\subsection{Non-Ideal Equation of State}

In addition to updated opacities, careful consideration must be given to
the equation of state (EOS) used in the calculation of very cool white dwarf atmospheres.
In stellar atmospheres, the EOS plays a dual role as it provides 1) the thermodynamic
properties of the material which affect the $(P,T)$ structure and the calculation of
the convective flux and 2) the relative abundances of all species which are
needed for the calculation of the opacity.
Saumon et al. (1994) and BSW used an ideal equation of state to compute the chemical
equilibrium between the various hydrogen species.  BSW found that for hydrogen white dwarf
models with $\Teff \ge 4000\,$K, the departures from a non-ideal EOS could
be neglected.  This is no longer true for the very cool white dwarfs we are considering 
here and we have used a fully non-ideal EOS in this calculation.  Our EOS is based on
the model of \cite{SCVH} who consider a mixture of H$_2$, H, H$^+$
and $e$ and account for the interactions between these species in a realistic
fashion.   In the regime of interest here ($P < 10^{12}\,$dyn/cm$^2$ and $T < 5000\,$K),
the molecular repulsion leads to an increased degree of dissociation of the molecules as
the pressure is raised.
By contrast, it is well known that an ideal EOS based on Saha equations  predicts 
complete {\it recombination}
at high pressures.  Recent experiments (\cite{Holmes95}) have revealed that the
EOS of Saumon, Chabrier \& Van Horn (1995)  underestimates the degree of molecular 
dissociation at high 
pressures.  By modifying the  interaction potentials used in the EOS calculation, we
obtain a new EOS which is in excellent agreement with all of the available high-pressure
data (\cite{SCWX98}).  The present atmosphere calculations include this updated
hydrogen EOS.

The non-ideal EOS does not include H$^-$, H$_3^+$, or H$_2^+$ which are important for
the calculation of the opacity.  These are introduced
as a perturbation  on the non-ideal EOS, where their abundances are calculated
from their respective Saha equation (\cite{LS92}).  Density effects on the 
internal partition functions
are described with an occupation probability formalism (\cite{HM88}).

The effects of the non-ideal EOS are threefold.  First, for given $P$ and $T$,  the density
is reduced compared to an ideal EOS.  This decreases the importance of H$_2$ CIA at depth in 
the models.  Second, the strong molecular
repulsion leads to a significant drop in the value of the adiabatic gradient (\cite{SCVH})
which affects the $(P,T)$ profile and the convective flux at depth. 
Finally, the pressure-dissociation of H$_2$ into H further reduces the
CIA contribution to the opacity and increases the H$^-$ contribution.

The effects of including non-ideal effects in the EOS are illustrated in Fig. 1
where models computed with the updated non-ideal EOS are compared with models
computed with an ideal EOS.  
The non-ideal effects are small for $\Teff=4000\,$K and increase for larger gravity and 
for  lower $\Teff$.
The decrease in  the adiabatic gradient for $\log P \wig> 10$ is readily apparent
in Fig. 1 since $\nabla_{\rm ad} \approx \nabla$ at depth.

The contours in Fig. 1 show how the non-ideal EOS results in a gradual decrease of 
the H$_2$ CIA opacity and an increase in the H$^-$ 
opacity as the pressure increases.  The lower CIA opacity results in a slightly
higher temperature in the radiative region of the model, {\it i.e.} the molecular
cooling of the radiative zone is reduced.    Triangles indicate the
location of the Rosseland mean photosphere ($\tau_{\sss R}=2/3$). Note, however,
that because the opacity is strongly non-gray,  the level of the ``photosphere''
varies greatly with wavelength.  For the $\Teff=2000\,$K, $\log g=8$ model, the
$\tau_\nu=2/3$  level varies over 2.5 order of magnitudes in pressure!

\section{The Colors of Very-Cool White Dwarfs}

The synthetic spectra and atmospheric models of very cool, pure hydrogen white dwarfs
are qualitatively similar to those obtained by Saumon et al. (1994)  for 
zero-metallicity brown dwarfs.
The main difference arises from the higher surface gravity of white dwarfs which enhances
the relative importance of the H$_2$ CIA opacity.
For $\Teff \wig> 4000\,$K, the
opacity is dominated by the relatively gray opacity of H$^-$, resulting in a spectrum
similar to that of a black body.  At lower $\Teff$, H$_2$ molecules become abundant in
the atmosphere and the H$_2$ CIA opacity becomes important.  The strong infrared bands
of CIA greatly reduce the infrared flux and force the flux to emerge at shorter wavelengths.
As initially reported by Saumon et al. (1994), the peak of the energy distribution of
atmospheres dominated by H$_2$ CIA opacity
shifts to the {\it blue} as $\Teff$ decreases.  The resulting SED is unlike
any that has been observed so far.  We find that in  the entire parameter range we have 
studied, very-cool white dwarfs are brightest in the $V$, $R$, and $I$ bands, and are
very faint in the $K$ band. Color-color diagrams are peculiar for most color 
combinations.  The colors $V-R$, $V-I$, $V-K$, $I-J$, $J-K$ and $H-K$ all reach a
maximum and then become {\it bluer} when $\Teff$ decreases below 4000$\,$K.

The colors derived from our SED's  ($\Teff \le 4000\,$K) are a very good match to those 
of BRL for $\Teff \ge 4000\,$K.  Small differences are caused by our use
of updated H$_2$ CIA cross-sections and of a non-ideal equation of state.
Since the H$_2$ opacity is strongest in the infrared, the decrease in CIA
and the increase in H$^-$ caused by the non-ideal effects of the EOS result in a redder spectrum than 
for a model based on an ideal EOS model ({\it e.g.} the models of BSW and BRL). 

From the three photometric bandpasses sampling the peak of the SED of very cool 
hydrogen-rich white 
dwarfs, we construct the  ($V-R$, $R-I$) two-color diagram shown in Fig. 2.  Our models
are shown along with the observed colors of a sample of cool hydrogen-rich white dwarfs (BRL).
The 
synthetic colors of the models naturally extend the  sequence of disk white dwarfs.
It is an interesting coincidence that the oldest, and therefore coolest, observed white
dwarfs lie just at the turnoff of the cooling sequence of Fig. 2.  This is also true
for other choices of colors.  The reddest hydrogen-rich white dwarfs therefore 
have $\Teff \approx
3800\,$K with $R-I \approx 0.7$ and $V-R \approx 0.7$.  Cooler white dwarfs become
very blue in this diagram.

White dwarfs with $\Teff \wig< 4000\,$K are very old and they cool at a constant 
radius.  By extrapolating the mass-radius relation of white dwarfs with carbon/oxygen
cores (\cite{Wood};\cite{Wood92}) we
can compute absolute magnitudes from our synthetic spectra.  Figure 3 shows our
grid of models in the ($M_V$, $V-I$) color-magnitude diagram.  The curves connecting
models with the same gravity are essentially identical to cooling curves.  The turnoff of the
white dwarf cooling sequence is located at $M_V \approx 17$ and $V-I\approx 1.4$.
This turnoff occurs just beyond the observed end of the disk white dwarf
sequence (\cite{LDM88}) and therefore corresponds to an age
of 10 to 12 Gyr ({\it e.g.} \cite{CSM}; \cite{H98}). Halo white dwarfs 
are thus
expected to have $V-I < 1.4$.   The long-dashed line in Fig. 3 shows the locus of models
with a gravity of $\log g=8$ computed with an ideal  EOS.  The non-ideal EOS significantly
affects the colors for $\Teff \le 2500\,$K or $V-I \le 0.6$.

\section{Conclusions}

We have calculated the most physically complete white dwarf atmosphere models
for pure hydrogen composition at very-low $\Teff$.  Elements central to this calculation are 
the use
of the latest opacities for hydrogen and the use of a non-ideal EOS.  Previous models
of very-cool atmospheres without metals (\cite{SB}; BSW; BRL) strongly
indicated that the onset of H$_2$ CIA opacity at low-$\Teff$ would dramatically affect
the colors of very cool white dwarfs.  We confirm that the colors of a  cooling
white dwarf  with a hydrogen-rich atmosphere reach a turning point and then become bluer 
as $\Teff$ continues to decrease,  a conclusion also reached by Hansen (1998). 

The colors we have calculated for very-cool white dwarfs allow for the 
unambiguous identification of white dwarfs in searches for the population
of MACHOs responsible for microlensing events in the direction of the LMC.  
Previous analyses of searches for halo white dwarfs can be reinterpreted in the
light of our quantitative prediction of the colors of halo white dwarfs.

Evolutionary calculations for old white dwarfs are in dire need of reliable model
atmospheres to serve as surface boundary conditions (\cite{LF1}; \cite{H98}).  It 
is the purpose
of our calculation to make white dwarfs better cosmic chronometers for ages above
10 Gyr.  This has immediate applications
in the calculation of globular clusters ages from their white dwarf
sequence (\cite{Rich97}), and in the determination of the age of the halo, if a significant
fraction of its mass is found to be  composed of
white dwarfs.   Since these estimates also provide an independent lower limit to the age 
of the universe, this method holds the promise of settling several important 
cosmological questions.

Finally, the low-resolution spectrum of an old globular cluster white dwarf would
be truly unique among stellar spectra, its SED being literally sculpted by the strong 
CIA of molecular hydrogen.

We thank P. Bergeron and G. Fontaine for useful discussions and A. Borysow for kindly
providing us with updated cross-sections for the collision-induced absorption by H$_2$.
This work was supported in part by NSF grant AST-9318970.

\newpage

\figcaption{Atmosphere model structures in the $P$ and $T$ plane.  Models
shown have $\Teff=$ 1500 to 3000$\,$K in steps of 500$\,$K (from bottom to
top) and a gravity of $g=10^8\,$cm/s$^2$.  The heavy long-dash curve divides the
diagram according to the dominant source of opacity.  Contours show the fractional change 
in the Rosseland mean opacity arising from the use of a non-ideal EOS, with dotted 
contours representing a {\it decrease} in 
$\kappa_{\sss R}$ (see text).  The contours levels are (from left to 
right): 2\%, 4\%, 6\%, 8\%,  and 10\%. The triangles indicate the level where
$\tau_{\sss R}=2/3$.} 

\figcaption{Color-color diagram for pure-hydrogen white dwarf atmospheres. Each heavy
curve represent a sequence of constant gravity models with $\log g=7.5$, 8, and 8.5
(from top to bottom).  Models with the same $\Teff$ (labeled) are 
connected by a thin line.
Solid dots represent the subset of stars observed by BRL with H-rich atmospheres and
known parallax.} 

\figcaption{Color-magnitude diagram for pure-hydrogen white dwarf atmospheres. 
See Fig. 2 for legend.  The long-dashed line shows the locus of models
computed with an ideal EOS.} 
\clearpage

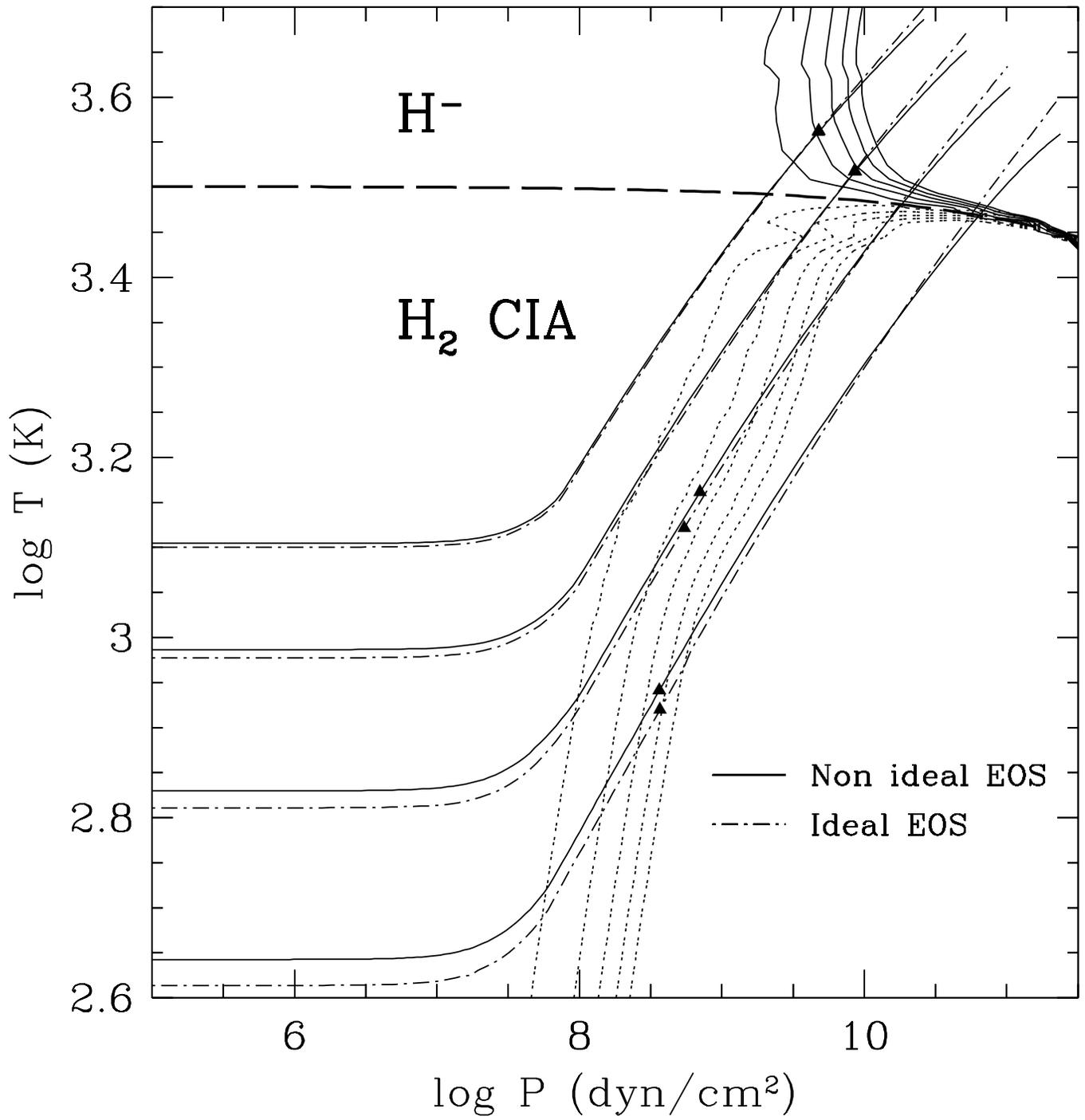
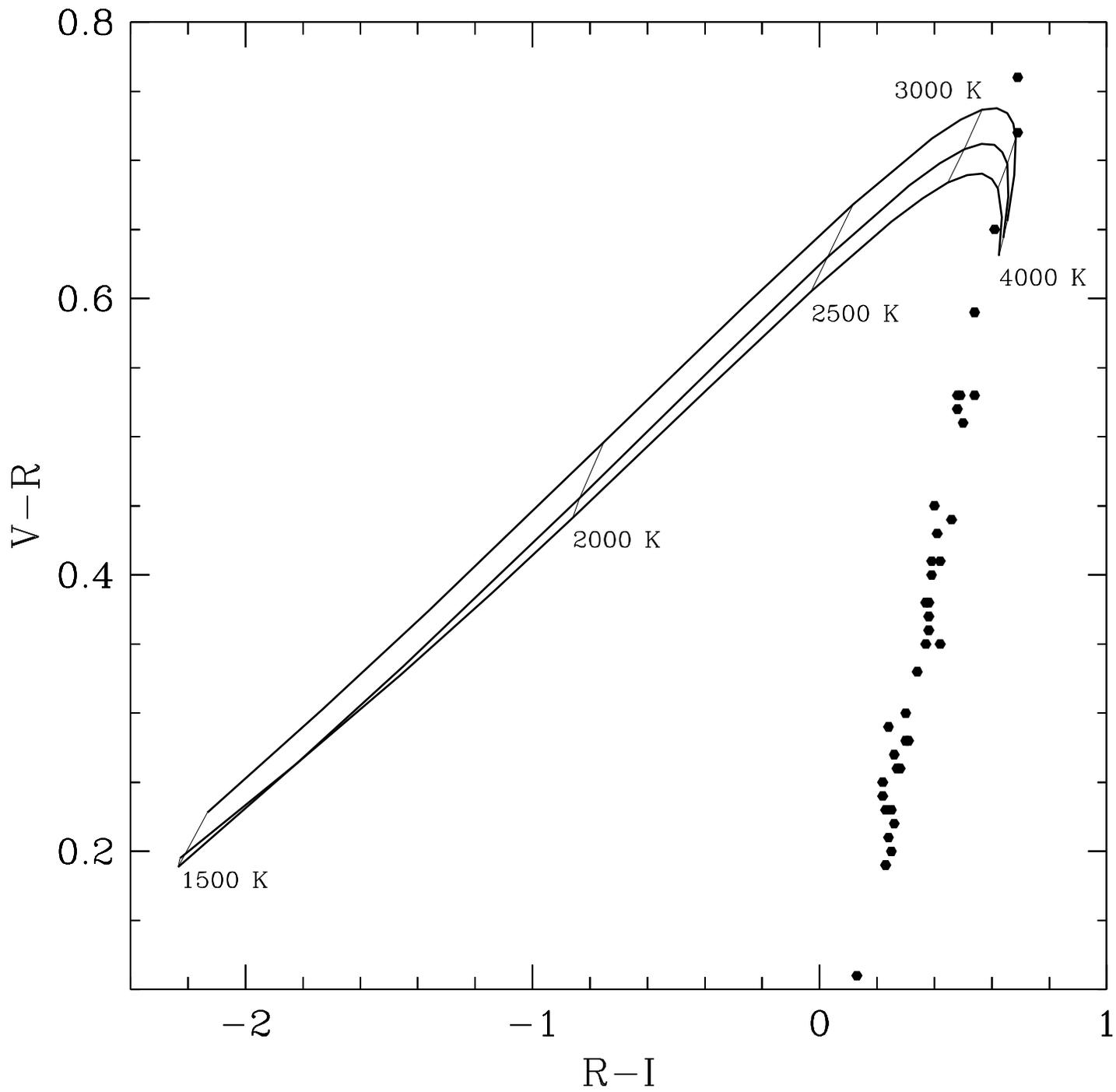
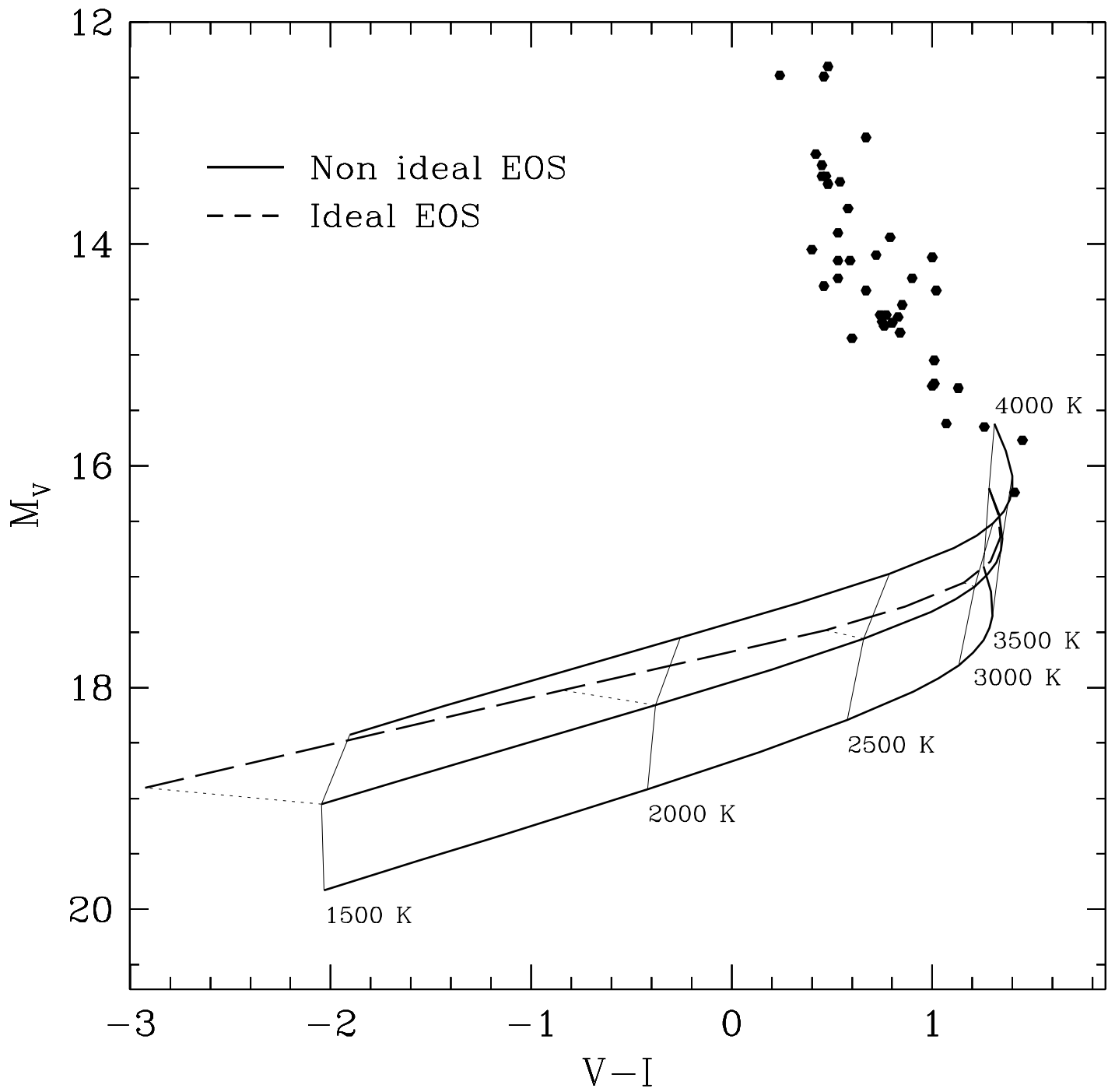
\end{document}